\documentclass[twocolumn,showpacs,preprintnumbers,amsmath,amssymb,prb,aps]{revtex4}

\usepackage{epsfig} 
\usepackage{graphicx}
\usepackage{dcolumn}
\usepackage{bm}
\usepackage{epstopdf}
\usepackage{amsmath}
\begin{document}

\title{Inverse photoemission spectroscopic studies on phase separated La$_{0.2}$Sr$_{0.8}$MnO$_{3}$}
\author{Navneet Singh$^1$}
\author{M. Maniraj$^2$}
\author{J. Nayak$^2$}
\author{S.K. Pandey$^3$}
\author{R. Bindu$^1$}
\altaffiliation{Corresponding author: bindu@iitmandi.ac.in}
\affiliation{$^1$School of Basic Sciences, Indian Institute of Technology Mandi, Kamand, Himachal Pradesh- 175005, India\linebreak
$^2$UGC-DAE Consortium for Scientific Research, Khandwa Road, Indore- 452001\linebreak
$^3$School of Engineering, Indian Institute of Technology Mandi, Kamand, Himachal Pradesh- 175005, India}

\date{\today}
\begin{abstract}
We have studied the temperature evolution of the inverse photoemission spectra of phase separated La$_{0.2}$Sr$_{0.8}$MnO$_{3}$. To identify the features in the room temperature experimental spectra, band structure calculations using Korringa-Kohn-Rostoker Green's function method were carried out. We find that the features generated by local moment disorder calculations give a better match with the experimental spectrum. In the insulating phase, we observed unusually an increased intensity at around the Fermi level. This puzzling behaviour is attributed to the shift in the chemical potential towards the conduction band.  The present results clearly show the importance of unoccupied electronic states in better understanding of the phase separated systems.

\end{abstract}


\maketitle

\section{Introduction}
In strongly correlated electron systems\cite{goodenough}, the subtle interplay between charge, spin, lattice and orbital degrees of freedom stabilize emergent phenomena like phase separation, colossal magneto resistance, superconductivity, multiferroicity, etc. The material under study, La$_{0.2}$Sr$_{0.8}$MnO$_{3}$ is one such kind of system, which reveals structural phase separation at the nanoscale\cite{PRBxrd}. At room temperature, this material stabilizes in a simple structure namely cubic. Like any other compounds in the Sr doped series, the compound under study does not lie in any of the phase boundary region thereby avoiding any complications due to the coexistence of phases lying in the phase boundary region\cite{hemberger}. Hence it is expected that this material provides a better platform to unravel its properties. This material undergoes structural, electronic and magnetic phase transitions; cubic to tetragonal, metal to insulator,  paramagnetic to C-type antiferromagnetic respectively, all at around Neel temperature (T$_{N}$ = 265 K)\cite{hemberger,Chmaissem,APL}.  Our results\cite{PRBPES,PRBxrd,NJP} based on temperature dependent transmission electron microscopy (TEM), synchrotron based x-ray diffraction (XRD), high resolution photoemission spectroscopy revealed inhomogenities at the nanoscale. To characterize such inhomogenities, it is important to obtain the behaviour of the chemical potential as a function of the thermodynamic variable.Photoemission spectroscopy is one such tool where one can get the information of the behaviour of chemical potential shift\cite{CP}.

In the case of photoemission spectroscopy, (i) the information of chemical potential shift is obtained based on the shift in the binding energy. There are several factors which contribute to the shift in the binding energy apart from the chemical potential shift \cite{CP}. These are chemical shift, madelung potential, screening potential, relaxation energy; (ii) There occurs ambiguity in extracting the binding energy shift if more than one unresolved features contribute to the core level. Such unresolved features occur due to various screening channels. So essentially, the information obtained about the chemical potential shift is rather indirect.

Inverse photoemission spectroscopy is one such technique which gives information of the unoccupied part of the electronic structure. It also gives a better insight into the behaviour of the chemical potential when compared with the results of the photoemission data, as no such complications will occur when extracting the information of the chemical potential shift. Until now, most of the work \cite{CP} carried out to extract chemical potential shift are based on the core level binding energy shift. In the systems which show electronic phase separation, the fractions of the coexisting phases vary in such a way that the chemical potential of both the phases remains the same. Keeping this in mind, it is important to extract this parameter unambiguously. We show here that this parameter can be extracted unambiguously using inverse photoemission spectroscopy. Apart from this, it is important to bear in mind that band structure calculations give information about the ground states.But the information provided by inverse photoemission spectroscopy is about the excited states. It has always been challenging to match the features generated by the excited states. The situation becomes more challenging, if the system is paramagnetic and belongs to the category of strongly correlated electron system. Here, we show that local moment disorder calculation gives a better representation of the features in the experimental inverse photoemission spectra. This motivated us to carry out temperature dependent inverse photoemission spectroscopic (IPES) studies on phase separated La$_{0.2}$Sr$_{0.8}$MnO$_{3}$.

\begin{figure}
 \vspace{-1ex}
\includegraphics [scale=0.4, angle=0]{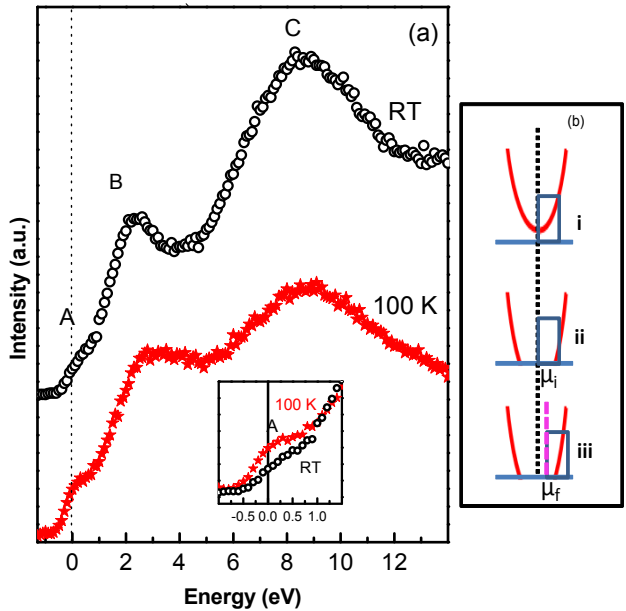}
 \vspace{-20ex}
\caption{(a) The inverse photoemission spectra collected at RT and 100 K. The inset shows the comparison of the spectra around the region of Fermi level, (b) Schematic displaying the shift in the chemical potential when the material undergoes transition from metal (panel (i)) to insulator. The panel (ii) shows the chemical potential, $\mu_{i}$ (black dot) in the insulating phase lying at the middle of the band gap. The panel(iii) shows the chemical potential, $\mu_{f}$ (magenta dash)  in the insulating phase shifted towards the conduction band.The rectangular block represents the region up to which the density of states contribute at the Fermi level due to instrumental broadening of 0.55 eV. }
\vspace{-2ex}
\end{figure}

\section{Experimental details}
Polycrystalline La$_{0.2}$Sr$_{0.8}$MnO$_{3}$ was prepared by conventional solid state route. Details of the sample preparation and characterisation are given elsewhere\cite{EJP}. To study the unoccupied density of states inverse photoemission experiments (IPES) were carried out on La$_{0.2}$Sr$_{0.8}$MnO$_{3}$ in ultra high vacuum at pressure 10$^{-11}$ mbar. The specimen was scrapped in vacuum using a diamond file in order to get rid of the surface contamination. An electron gun of Stoffel Jhnson\cite{JPE,Maniraj} design and an acetone filled band pass detector with CaF$_{2}$ window is used for the experiments.\cite{RSI,CS} The experiments were carried out in the isochromat mode where the energy of the incident electron was varied in steps of 0.1 eV and photons of fixed energy (9.9 eV) were detected with an overall instrumental resolution of 0.55 eV. The electron beam current variation as a function of kinetic energy was accounted for by normalizing the measured counts by the sample current at each step, as in our previous studies.\cite{IPES}

\section{Computational details}
The non-magnetic, ferromagnetic and local moment disorder (LMD)\cite{H.Akai} calculations for La$_{0.2}$Sr$_{0.8}$MnO$_{3}$ were carried out by using Korringa-Kohn-Rostoker (KKR) Green's function method.\cite{http} In the compound under study 80\% Sr is doped at La site. These doping effects were studied well under the coherent potential approximation (CPA). As the crystal structure is cubic, lattice parameter used for calculation are a=7.18886 Bohr. Muffin tin radii for La/Sr, Mn and O  3.064 Bohr, 1.566 Bohr and 2.05 Bohr, respectively are used for the calculations. The exchange correlation function used for the calculation were taken after Vosko, Wilk and Nusair.\cite{vosko} The self consistency was achieved by demanding convergence of total energy to be less than 10$^{-6}$ Ryd/cell.

\section{Results and discussions}
Room temperature (RT) inverse photoemission spectrum for La$_{0.2}$Sr$_{0.8}$MnO$_{3}$ is shown in Fig. 1a.The Fermi level ($\epsilon$$_{F}$) is marked as E = 0 eV. In this spectrum, there are three clear features A, B and C observed at and above $\varepsilon$$_{F}$ . A small kink upto around 0.7 eV is labeled as A and two broad features are labeled as B and C. Peak B is centered around 2.3 eV and spread up to about 4.5 eV, whereas the peak C is centered around 8.8 eV and spread upto 13 eV. To identify these features band structure calculations were carried out.

\begin{figure}
\vspace{-2ex}
\includegraphics [scale=0.4, angle=0]{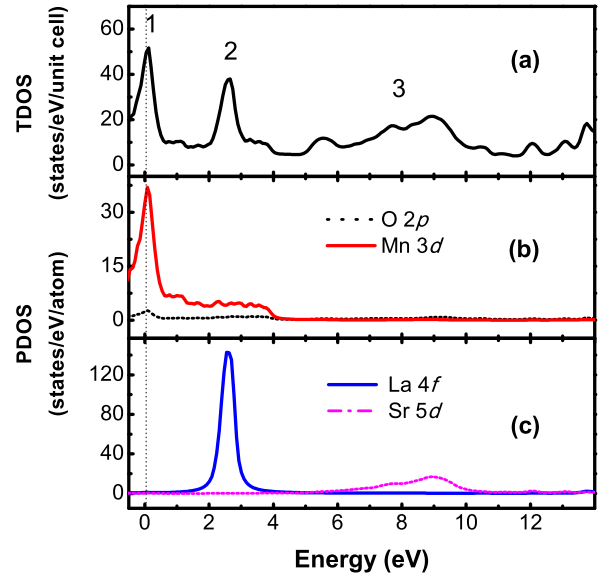}
\vspace{-15ex}
\caption{(a) The total density of states (TDOS) obtained based on non magnetic calculations, (b) The black dot and red line represent O 2$\emph{p}$ and Mn 3$\emph{d}$ partial density of states (PDOS), respectively and (c) The blue line and magenta dash dot represent La 4\emph{f} and Sr 5\emph{d} PDOS, respectively.}
\vspace{-2ex}
\end{figure}

\begin{figure}
\vspace{-2ex}
\includegraphics [scale=0.4, angle=0]{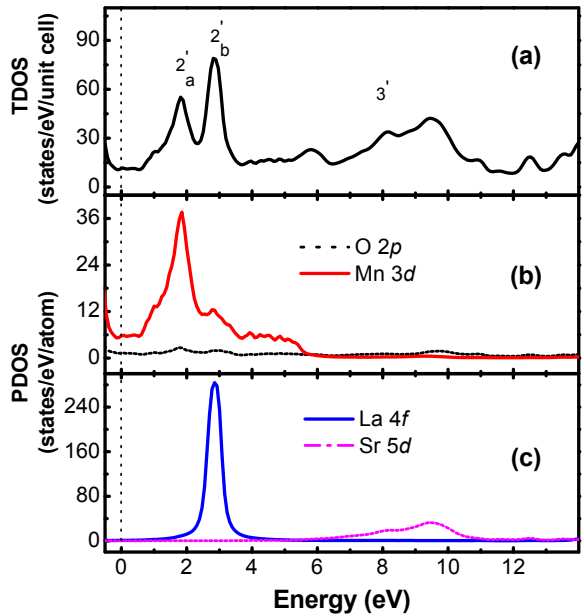}
\vspace{-20ex}
\caption{(a) The total density of states (TDOS) obtained based on ferromagnetic calculations, (b) The black dot and red line represent O 2$\emph{p}$ and Mn 3$\emph{d}$ partial density of states (PDOS), respectively and (c) The blue line and magenta dash dot represent La 4\emph{f} and Sr 5\emph{d} PDOS, respectively.}
\vspace{-2ex}
\end{figure}

To begin with, we performed non magnetic calculations to understand the paramagnetic phase displayed by this compound at RT. The results indicate three important features in the total density of states (TDOS), Fig.2a. The feature 1 which is close to the $\varepsilon$$_{F}$ is centered around 0.1 eV; feature 2 is spread in the region 2 to 3 eV and the third one in the broad region from 5 to 11 eV. The partial density of states as displayed in Figs.2b \& c indicate that the feature 1 corresponds mainly to Mn 3\emph{d} states with negligible contribution from O 2\emph{p} states; the feature 2 corresponds to La 4\emph{f} and the feature 3 to mainly Sr 5\emph{d} states. On comparing the calculated DOS with experimental spectrum we observe that only features 2 and 3 are fairly matching with the peaks B and C in the experimental spectrum. The discrepancy in the matching of feature 1 with peak A is mainly due to the non-magnetic calculations which are based on the itinerant model of paramagnetism. In the itinerant model of paramagnetism, no Hund's like on-site exchange interaction at magnetic ion (Mn) site are taken into account. But for the material under study it is important to invoke the contribution of on-site exchange interaction as this belongs to strongly correlated electron systems where the Mn 3$\emph{d}$ electrons are highly localised.

\begin{figure}
\vspace{-2ex}
\includegraphics [scale=0.4, angle=0]{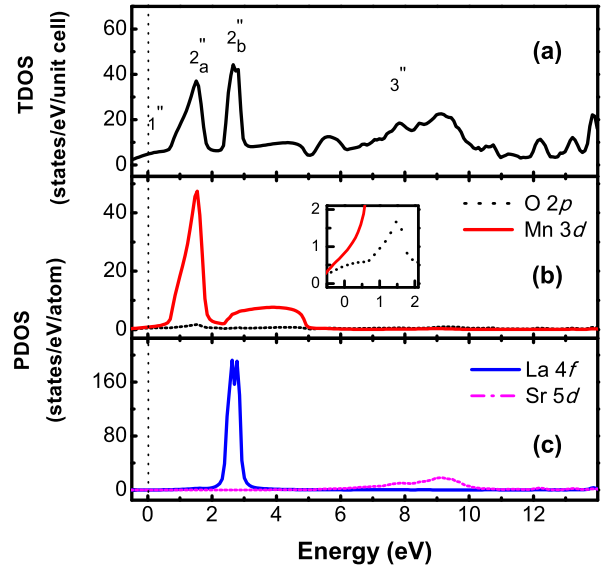}
\vspace{-17ex}
\caption{(a) Total density of states (TDOS) obtained based on LMD calculations, (b) The black dot and red line represent O 2$\emph{p}$ and Mn 3$\emph{d}$ partial density of states (PDOS), respectively. The inset shows the contribution from O 2\emph{p}, Mn 3\emph{d} PDOS close to the Fermi level and (c) The blue line and magenta dash dot represent La 4\emph{f} and Sr 5\emph{d} PDOS, respectively.}
\vspace{-2ex}
\end{figure}

As a next step; to capture the localized picture of 3$\emph{d}$-electrons and to include the exchange interactions, we performed ferromagnetic calculations, Fig.3. In the TDOS, we observe three main features labeled as 2$'$$_{a}$,2$'$$_{b}$ and 3$'$. The features 2$'$$_{a}$\&2$'$$_{b}$ are extended in the energy range 0.5 to 4 eV and the feature 3$'$ is spread in the region from 6 to 11 eV. The feature 2$'$$_{a}$ has strong contribution from Mn  3$\emph{d}$ and weak contribution from O  2$\emph{p}$ states. The feature 2$'$$_{b}$ has main contribution from La 4$\emph{f}$ states with small contribution from Mn 3$\emph{d}$ states. The feature 3$'$ is mainly contributed by Sr 5$\emph{d}$ states and negligible contribution from O 2$\emph{p}$ states. On comparing the calculated DOS with the experimental spectrum we observe, only peak B matches fairly good with features 2$'$$_{a}$ and 2$'$$_{b}$ and peak C with feature 3$'$.  No feature corresponding to peak A has been generated in the calculated DOS. This discrepancy could be due to the following reason. In the ferromagnetic calculations, there are contributions from both on-site and inter-site exchange interactions. The former interaction gives rise to local magnetic moment while the latter gives rise to long range magnetic order. The compound under study is paramagnetic at room temperature, so the missing of the feature corresponding to peak A could be due to the long range magnetic ordering generated by the calculation.

To capture the localized picture of  \emph{d}-electrons without inter-site exchange interactions, local moment disorder (LMD) calculations were carried out, Fig.4. We observe four features 1$''$, 2$''$$_{a}$,2$''$$_{b}$ and 3$''$  in the calculated TDOS.  The feature 1$''$ is observed as small kink upto 0.5 eV, features 2$''$$_{a}$\&2$''$$_{b}$ are extended in the energy range 0.6 to 5 eV and the feature 3$''$ is spread in the region from 6 to 11 eV. On comparing the calculated DOS with the experimental spectrum, we attribute feature 1$''$  to Mn \emph{3d} and O \emph{2p} hybridized states; the features 2$''$$_{a}$ and 2$''$$_{b}$ to Mn \emph{3d} and La \emph{4f} states, respectively and the feature 3$''$ mainly to Sr \emph{5d} states. The features generated in the calculations are matching fairly good with peaks A, B and C of experimental spectrum.

Having identified the features in the room temperature spectrum, we now discuss the results of the temperature dependent spectra displayed in Fig.1a. On reducing the temperature to 100 K, we find an increased intensity upto about 0.9 eV above $\varepsilon$$_{F}$ as shown in the inset of the figure. The studies on temperature dependent high resolution photoemission spectra \cite{PRBPES} have shown that hard gap is opened up only at 200 K even though metal to insulator transition found from the resistivity measurements occurs at 265 K. In this situation, it is normally expected that for the inverse photoemission spectra, the intensity at around $\varepsilon$$_{F}$ should also decrease. But we observe opposite behaviour which is puzzling. In the first instance, several possibilities come into the mind when one observes such increase in the intensity. The possibilities could be either due to (a) charging effects or (b) destruction of insulating state as a result of electron irradiation\cite{M. Hervieu} or (c) current induced phase transition at the surface or (d) chemical potential shift towards the conduction band.

After careful analysis of the data we rule out the first possibility due to the following reasons; (i)At 100 K, the charging effect was not observed even in the X-ray photoemission spectra \cite{PRBPES}; (ii)Had there been charging then one would have expected the spectrum to be shifted towards higher energy. So in this case one would expect depletion of intensity at the Fermi level which is not the case we observe; (iii) Apart from this, we also observe the position of the higher energy peaks at RT and 100 K i.e. around 3 and 8 eV to be almost the same despite the material are in two different electronic phases namely room temperature metallic and low temperature (100 K) insulating phase.	 .

The second possibility can be ruled out based on our earlier high resolution XRD and TEM results \cite{PRBxrd}, \cite{PRBPES}. This compound exhibits co-existence of charge ordered and twinned phase in a wide temperature range (even at 300 K) which becomes more prominent in the temperature range 260 K to 200 K. With decrease in temperature, it is found that the fractions of the twinned phase increase at the cost of the charge ordered phase and at 100 K there is about 10 \% of charge ordered phase. In the light of the above fact, there is no question of melting of charge ordering at 100 K due to irradiation as the electron energy used in TEM is more than 100 times higher than that used in the IPES experiments. Had there been melting of charge ordering due to electron irradiation, we would not have observed robust charge ordered phase in the TEM experiments. If we consider the third possibility, then the intensity at $\varepsilon$$_{F}$ for 300 K will be more than that in 100 K which is opposite to the experimental observation. Thus one can also rule out the third possibility.

Regarding the fourth possibility, it is important to recollect our photoemission results\cite{NJP,PRBPES}. The signature of above phase separation was also confirmed based on the behaviour of the chemical potential shift calculated from the binding energy shift of the core level spectra. Even though the metal to insulator transition occurred at around 265 K, the chemical potential shift towards higher binding energy was observed below 200 K. The pinning of the chemical potential shift in the temperature range 300 to 200 K indicates electronic phase separation. In the region of phase separation, pseudo gap was observed and on further reducing the temperature hard gap was opened in the antiferromagnetic phase. The observed shift in the binding energy in the PES data has been explained in terms of shift in the chemical potential towards the conduction band\cite{PRBPES}. The observed increment in the intensity of peak A in the IPES data at 100 K can also be understood based on the shift in the chemical potential towards the conduction band as explained below.

The pseudogap was observed in the phase separated region, we assume the behaviour of the spectra close to $\varepsilon$$_{F}$ (both in the occupied and unoccupied region) to a parabola. So as the compound enters the insulating phase, states at $\varepsilon$$_{F}$ undergo depletion. In Fig. 1b, the states contributing to the intensity at $\varepsilon$$_{F}$ due to the instrumental broadening (0.55 eV) is represented by the region marked by the rectangular block. The panel (i) shows finite density of states at $\varepsilon$$_{F}$ corresponding to RT metallic phase. Here, the states within the rectangular block are expected to contribute at $\varepsilon$$_{F}$ which leads to finite intensity at $\varepsilon$$_{F}$. As the material enters into the insulating phase, the hard gap is opened. The states in the region marked by the rectangular block are expected to contribute at $\varepsilon$$_{F}$, panel (ii). So in the IPES data one expects decrement in the intensity at $\varepsilon$$_{F}$ with the chemical potential lying in the middle of the gap. Such behaviour is not in line with the experimental result, inset of Fig 1a. We now discuss the situation where the chemical potential is shifted towards the conduction band, panel (iii). When such shift is taken into account, the contribution of the states belonging to the rectangular block is expected to be more at $\varepsilon$$_{F}$. This leads to the increase in the intensity of peak A with decrease in temperature.  It is important to point out that, in the ultraviolet photoemission spectra\cite{NJP,PRBPES}, it is essentially because of the high resolution (5 meV) of the photoemission set up, a clear gap was observed at low temperature. The band gap at 130 K using PES was estimated to be of about 150 meV, under the assumption that there is no chemical potential shift. But as we have observed chemical potential shift at 100 K, the band gap is expected to lie between 75 to 150 meV.

\section{Conclusions}
Inverse photoemission spectra were collected on La$_{0.2}$Sr$_{0.8}$MnO$_{3}$ at 300 and 100 K. The features in the inverse photoemission spectra were identified using band structure calculations and it was found that local moment disorder calculation gives a better match with the experimental spectra. As the compound becomes insulating, at around $\epsilon$$_{F}$, an increase in the intensity was found. This behaviour is in contrast with the results of the high resolution photoemission spectra. After detailed analysis, we find that this unusual behaviour can be understood based on the shift in the chemical potential. The present work clearly suggests the importance of occupied and unoccupied electronic states to understand the behaviour of systems with electronic phase separation.

\section{Acknowledgement}

The authors thank Dr. S.R. Barman, UGC-CSR, Indore for fruitful discussions.

\end{document}